\documentclass[traditabstract]{aa} 
\usepackage{graphicx}
\usepackage{txfonts}
\usepackage{array}
\usepackage{longtable}
\usepackage{multirow}

\usepackage{natbib}

\begin{document}
   \title{Known Galactic field Blazhko stars}


   \author{M. Skarka
          \inst{1}
          }

   \institute{Department of Theoretical Physics and Astrophysics, Faculty of science, Masaryk University, 
   						Kotl\'a\v rsk\'a 2, Brno, Czech Republic\\
              \email{maska@physics.muni.cz}
             }




  \abstract
  {A list of known Galactic field stars exhibiting Blazhko effect containing 242 stars is presented. All the entries including their designations, positions, pulsation, and Blazhko periods were collected from the available literature. The actual values of parameters are given.}
  {}{}{}{}
   \keywords{Catalogs -- horizontal-branch -- RR Lyr}

   \maketitle
%

\section{Introduction}

In the era of automatic sky surveys and space telescopes, it is increasingly 
obvious that there is a high percentage of RR~Lyrae type 
stars exhibiting the Blazhko effect \citep[more than 40~\%,][]{kolenberg2010}. 
The light curves of such stars are amplitude- or phase-modulated 
with the periods typically in the order of tens to hundreds of days\footnote{
There are also several stars with changing Blazhko effect}. This behaviour 
is named after one of its discoverers, S. N. \citet{blazhko1907}, who noticed 
it in RW Dra. 

Many theories, such as resonances between radial and nonradial modes 
\citep{dziembowski2004}, effects connected with convection \citep{stothers2006}, 
etc, have tried to describe observed properties, but the 
explanation of the nature of the Blazhko effect is still missing. 
A brief overview of the Blazhko effect can be found in \citet{kovacs2009} and \citet{kolenberg2012}. 
Nowadays, the most likely explanation of the Blazhko effect is considered 
resonance between the fundamental radial mode and the ninth overtone 
\citep{buchler2011}.

Some partial lists of the Blazhko stars have been published 
\citep[e.g.][]{smith1995, sodor2005, leborgne2012}, but an overall list has 
not been available until now. Such a list allows observers and all 
interested astronomers to quickly check the Blazhko instability of 
the star in one place and see whether the star needs further 
observations or not.

\section{The list}

The list contains values based on the data from sky surveys ASAS 
\citep{szczygiel2007} and NSVS \citep{wils2006}, as well as the data 
based on {\it O-C} diagram analyses \citep{leborgne2012}, and finally 
it contains values obtained in detailed studies of many of these stars. 
Stars with only one modulation period are listed in Table 1, stars 
with multiple Blazhko period are in Table 2, and RR Lyraes with variable 
Blazhko period are in Table 3.

Coordinates and magnitude ranges were taken from the VSX database \citep{vsx}. 
If there is more than one available value of the Blazhko period, 
then the value with the highest priority is given. 
The hierarchy of cited references is the following: Values taken from 
detail studies have the highest priority, data from surveys have 
lower priority, and the the data based on {\it O-C} studies the lowest priority. 
There are some exceptions, mainly if the data 
with higher priority were published before 1990 or if the data are 
of worse quality than values with lower priority. If the values of 
Blazhko periods differ more than one day, then all available values are 
listed.

The stars are sorted by right ascension. If possible, 
the ASAS and other designations of the stars were transformed to 
GCVS names. This is the case of V1820 Ori, BB Lep, V339 Lup, MR Lib, 
V559 Hya, V552 Hya, V701 Pup, LR Eri, IY Eri, GW Cet, DZ Oct, V354 Vir, 
V419 Vir, V476 Vir, V551 Vir, OR Com, BT Sco, V1319 Sco, BT Ant, 
AD UMa, NS UMa, PP UMa, KV Cnc, AI Crt, and FR Psc, which were noted in \citet{wils2006} 
and \citet{szczygiel2007} in other forms.

There are three stars with special characteristics in the list. These objecst 
deserve to be observed as a matter of priority. Some indications show that 
BV Aqr is a RRd type \citep{jerzykiewicz1995}. VX Her is suspected to be a member 
of an eclipsing binary star \citep{fitch1966}, which is the first possible occurrence
of such objects among RR Lyraes. SU Col probably has three modulation periods, which 
is also unique behaviour \citep{szczygiel2007}. Some stars have a very 
short period (less than 0.23 d) or amplitude (about 0.1). These objects should be 
observed first.

Objects with very long (more than 1000 d) Blazko period are marked by 
a colon in the second column. The same mark may be used for that are only suspected 
of the Blazhko effect or whose Blazhko period is not well determined. Regularly updated list 
with Tables 1--3 is also available on the web page http://physics.muni.cz/$\sim$blasgalf.





\begin{acknowledgements}
Work on the paper have been supported by GACR project GD205/08/H005, MU~MUNI/A/0968/2009. 
The International Variable Star Index (VSX) database, operated at AAVSO, 
Cambridge, Massachusetts, USA have been used. This research also have made 
use of VizieR catalogue access tool, CDS, Strasbourg, France. I would like to 
thank Miloslav Zejda for usefull comments and suggestions.
\end{acknowledgements}

\onecolumn

\def\arraystretch{1.3}
\begin{footnotesize}
\setlength\tabcolsep{1.1ex}

\begin{longtable}{lccllrrcrrr}

\caption{List of Blazhko stars with one modulation period}\\
\hline 
\hline

\centering

		Star~~~~~~		&	&	RA					&	~~~~~~DE		&	Type	&	V$_{\mathrm{max}}$	&	V$_{\mathrm{min}}$ 	&	P$_{\mathrm{puls}}$	&	Ref.	&	~~P$_{\mathrm{Bl}}$~~ 	&	Ref.	\\
						&	&							&							&				&	[mag]								&	[mag]								&	[d]									&				&	~[d]~~									&				\\
\hline

\endfirsthead
\caption{continued.}\\
\hline\hline
		Star~~~~~~		&	&	RA					&	~~~~~~DE		&	Type	&	V$_{\mathrm{max}}$	&	V$_{\mathrm{min}}$ 	&	P$_{\mathrm{puls}}$	&	Ref.	&	~~P$_{\mathrm{Bl}}$~~ 	&	Ref.	\\
						&	&							&							&				&	[mag]								&	[mag]								&	[d]									&				&	~[d]~~									&				\\
\hline
\endhead
\hline
\endfoot

RY Psc						&		&	00	11	41.10	&	-01	44	55.3	&	RRab	&	11.82	&	12.72	&	0.5297456	&	25	&	154.53	&	25\\
OV And						&	:	&	00	20	44.86	&	40	49	41.8	&	RRab	&	10.90	&	11.26	&	0.47060		&	3		&	27	&	27\\
SW And						&	:	&	00	23	43.09	&	29	24	03.6	&	RRab	&	9.14	&	10.09	&	0.4422618	&	3		&	36.8	&	26\\
RX Cet						&		&	00	33	38.28	&	-15	29	14.9	&	RRab	&	11.01	&	11.75	&	0.5736918	&	18	&	255.5	&	25\\
ASAS003514-0415.0	&	:	&	00	35	14.00	&	-04	15	00.0	&	RRc		&	12.92	&	13.73	&	0.3445751	&	25	&	1616.29	&	25\\
ASAS003706-4317.7	&		&	00	37	06.00	&	-43	17	42.0	&	RRab	&	13.29	&	14.60	&	0.6275343	&	25	&	187.12	&	25\\
SW Psc						&	:	&	00	41	19.41	&	05	20	47.0	&	RRab	&	13.3	&	14.9	&	0.521265	&	3		&	34.5	&	26\\
FR Psc						&	:	&	00	47	57.06	&	11	42	43.5	&	RRab	&	11.5	&	12.8	&	0.45568		&	3		&	55	&	27\\
RU Cet						&	:	&	01	00	40.30	&	-15	57	27.6	&	RRab	&	11.10	&	12.03	&	0.5862844	&	3		&	98	&	26\\
ET Cep						&		&	01	02	23.30	&	85	23	49.2	&	RRab	&	13.5	&	14.5	&	0.49716		&	3		&	...	&	37\\
DR And						&	:	&	01	05	10.71	&	34	13	06.3	&	RRab	&	12.03	&	13.15	&	0.563118	&	15	&	$\sim$57.5	&	15\\
CS Phe						&		&	01	09	49.45	&	-44	18	53.5	&	RRab	&	12.66	&	13.53	&	0.4843964	&	3		&	62.5	&	17\\
RU Psc						&	:	&	01	14	26.04	&	24	24	56.4	&	RRc		&	9.93	&	10.40	&	0.390385	&	3		&	28	&	26\\
AM Tuc						&	:	&	01	18	30.65	&	-67	55	05.0	&	RRc		&	11.39	&	11.87	&	0.4057948	&	25	&	1748.86	&	25\\
XY And						&		&	01	26	42.41	&	34	04	07.4	&	RRab	&	12.90	&	14.22	&	0.3987247	&	9		&	41.37	&	9\\
GW Cet						&		&	01	28	48.27	&	-11	27	12.6	&	RRab	&	12.4	&	13.9	&	0.516648	&	3		&	84.99	&	25\\
ASAS013140-4957.3	&		&	01	31	40.60	&	-49	57	18.9	&	RRab	&	12.12	&	13.22	&	0.464329	&	25	&	40.17	&	25\\
UX Tri						&		&	01	45	35.01	&	31	22	49.6	&	RRab	&	13.07	&	14.50	&	0.4669218	&	13	&	43.7	&	13\\
IY Eri						&	:	&	02	07	28.18	&	-57	52	09.5	&	RRc		&	10.86	&	11.31	&	0.3750261	&	25	&	1673.36	&	25\\
SS For						&		&	02	07	51.98	&	-26	51	57.7	&	RRab	&	9.45	&	10.60	&	0.495433	&	19	&	34.94	&	19\\
RV Cet						&		&	02	15	14.90	&	-10	48	00.7	&	RRab	&	10.35	&	11.22	&	0.6234139	&	25	&	112.05	&	25\\
ASAS022637-4119.7	&		&	02	26	37.00	&	-41	19	42.0	&	RRc		&	10.08	&	10.21	&	0.2941932	&	25	&	357.94	&	25\\
RV Hor						&		&	02	50	20.47	&	-64	15	40.9	&	RRab	&	12.9	&	14.5	&	0.5724975	&	25	&	79.81	&	25\\
ASAS030534-3116.1	&		&	03	05	34.00	&	-31	16	06.0	&	RRab	&	12.53	&	14.00	&	0.4964538	&	25	&	6.77	&	25\\
RX For						&		&	03	11	13.22	&	-26	28	58.8	&	RRab	&	11.12	&	12.46	&	0.59731		&	18	&	31.79	&	23\\
ASAS031408-3446.4	&	:	&	03	14	08.00	&	-34	46	24.0	&	RRc		&	11.54	&	12.07	&	0.3124235	&	25	&	1241.77	&	25\\
ASAS032438-2334.7	&		&	03	24	38.00	&	-23	34	42.0	&	RRab	&	12.09	&	13.03	&	0.6296339	&	25	&	335.29	&	25\\
X Ret							&		&	03	25	20.10	&	-65	03	18.6	&	RRab	&	11.16	&	12.14	&	0.4920082	&	25	&	160.64	&	25\\
ASAS033108+0713.4	&	:	&	03	31	08.38	&	07	13	24.9	&	RRab	&	10.74	&	10.87	&	0.5281963	&	25	&	1442.79	&	25\\
LR Eri						&		&	04	00	10.76	&	-19	49	37.1	&	RRab	&	12.0	&	13.0	&	0.60225		&	17	&	122	&	17\\
FM Per						&	:	&	04	03	27.06	&	47	59	51.7	&	RRab	&	12.02	&	13.24	&	0.489256	&	14	&	$\sim$122	&	14\\
									&		&								&								&				&				&				&						&			&	20	&	27\\
AH Cam						&		&	04	06	38.89	&	55	29	59.7	&	RRab	&	11.31	&	12.33	&	0.3687346	&	3		&	10.83	&	23\\
XY Eri						&		&	04	11	16.78	&	-13	50	54.3	&	RRab	&	12.39	&	13.27	&	0.55426		&	18	&	50.2	&	23\\
BR Tau						&		&	04	34	42.91	&	21	46	21.7	&	RRab	&	12.07	&	13.38	&	0.3905928	&	8		&	19.3	&	8\\
AL Pic						&		&	04	41	30.80	&	-52	16	37.0	&	RRab	&	12.8	&	14.0	&	0.54861		&	18	&	34	&	17\\
U  Cae						&		&	04	53	14.41	&	-37	49	15.9	&	RRab	&	11.43	&	12.70	&	0.4197835	&	3		&	22.8	&	23\\
NSV1856						&		&	05	08	38.65	&	-56	02	57.5	&	RRab	&	12.5	&	13.5	&	0.5160761	&	25	&	786.91	&	25\\
RY Col						&		&	05	15	07.78	&	-41	37	41.7	&	RRab	&	10.44	&	11.24	&	0.4788368	&	25	&	82.08	&	23\\
ASAS052402-2247.4	&		&	05	24	02.00	&	-22	47	24.0	&	RRab	&	13.40	&	14.59	&	0.6498517	&	25	&	10.23	&	25\\
ASAS052840-5316.2	&		&	05	28	40.00	&	-53	16	12.0	&	RRc		&	13.38	&	14.05	&	0.3678062	&	25	&	453.43	&	25\\
ASAS053022-3234.8	&		&	05	30	22.00	&	-32	34	48.0	&	RRc		&	11.77	&	11.97	&	0.2331114	&	25	&	364.6	&	25\\
ASAS053628-3837.0	&	:	&	05	36	28.00	&	-38	37	00.0	&	RRc		&	12.78	&	13.23	&	0.3714727	&	25	&	1394.31	&	25\\
BB Lep						&		&	05	42	30.18	&	-16	22	54.5	&	RRab	&	11.85	&	12.71	&	0.5389135	&	25	&	22.84	&	25\\
ASAS054843-1627.0	&	:	&	05	48	43.00	&	-16	27	00.0	&	RRab	&	12.96	&	13.57	&	0.3767273	&	25	&	1663.06	&	25\\
ASAS055322-5417.9	&		&	05	53	22.00	&	-54	17	54.0	&	RRc		&	12.82	&	13.38	&	0.2452638	&	25	&	381.58	&	25\\
V1820 Ori					&		&	05	54	37.13	&	04	54	11.4	&	RRab	&	12.5	&	13.4	&	0.47927		&	3		&	28	&	27\\
VW Dor						&		&	06	07	45.71	&	-66	58	38.8	&	RRab	&	11.1	&	12.25	&	0.57057		&	18	&	25.99	&	23\\
RX Col						&		&	06	13	14.74	&	-37	15	00.6	&	RRab	&	12.32	&	12.95	&	0.59376		&	18	&	130	&	17\\
									&		&								&								&				&				&				&						&			&	137.77	&	23\\
ST Pic						&		&	06	14	01.17	&	-61	28	23.5	&	RRab	&	9.29	&	9.77	&	0.4857445	&	25	&	117.9	&	25\\
ASAS062326+0005.8	&		&	06	23	26.00	&	00	05	48.0	&	RRab	&	11.96	&	12.59	&	0.551317	&	25	&	37.24	&	25\\
ASAS064615-4319.2	&	:	&	06	46	15.00	&	-43	19	12.0	&	RRc		&	12.93	&	13.63	&	0.3186021	&	25	&	1639.61	&	25\\
ASAS070001-3732.5	&		&	07	00	00.65	&	-37	32	31.5	&	RRab	&	11.85	&	12.74	&	0.4941843	&	25	&	116.96	&	25\\
ASAS070854+1919.7	&		&	07	08	53.85	&	19	19	38.0	&	RRab	&	8.69	&	8.79	&	0.7789192	&	25	&	7.92	&	25\\
ASAS071549-4405.3	&		&	07	15	49.00	&	-44	05	18.0	&	RRc		&	13.04	&	13.67	&	0.3144609	&	25	&	488.4	&	25\\
RR Gem						&		&	07	21	33.53	&	30	52	59.5	&	RRab	&	10.62	&	11.99	&	0.3972884	&	4		&	7.23	&	4\\
ASAS080249-5913.5	&	:	&	08	02	48.94	&	-59	13	28.0	&	RRc		&	11.75	&	12.00	&	0.3541891	&	25	&	1185.26	&	25\\
ASAS080318-2530.1	&	:	&	08	03	18.25	&	-25	30	06.3	&	RRc		&	12.5	&	13.0	&	0.2697723	&	25	&	1542.02	&	25\\
SS Cnc						&		&	08	06	25.59	&	23	15	05.7	&	RRab	&	11.49	&	12.72	&	0.367337	&	2		&	5.309	&	2\\
DD Hya						&		&	08	12	31.81	&	02	50	05.0	&	RRab	&	11.57	&	12.71	&	0.501776	&	3		&	34.59	&	23\\
V701 Pup					&		&	08	19	32.86	&	-23	58	09.8	&	RRc		&	10.45	&	10.75	&	0.2856671	&	25	&	8.1	&	25\\
NS UMa						&	:	&	08	24	24.73	&	65	43	03.4	&	RRab	&	10.75	&	11.35	&	0.59910		&	3		&	65	&	27\\
TT Cnc						&		&	08	32	55.18	&	13	11	28.5	&	RRab	&	10.93	&	11.57	&	0.56340		&	23	&	89.02	&	23\\
KV Cnc						&	:	&	08	40	02.42	&	27	43	31.6	&	RRab	&	11.9	&	13.0	&	0.50200		&	3		&	42	&	27\\
SV Vol						&		&	08	48	32.64	&	-71	39	14.8	&	RRab	&	11.78	&	12.57	&	0.6099118	&	3		&	85.47	&	23\\
PP UMa						&	:	&	08	52	15.07	&	70	26	23.9	&	RRab	&	13.45	&	14.35	&	0.51869	  &	3		&	46	&	27\\
ASAS085254-0300.3	&		&	08	52	54.00	&	-03	00	18.0	&	RRc		&	12.42	&	12.65	&	0.2669022	&	25	&	11.8	&	25\\
DZ Oct						&		&	08	54	48.33	&	-83	16	57.2	&	RRab	&	12.2	&	13.5	&	0.47786		&	18	&	36.8	&	25\\
ASAS090900-0410.4	&		&	09	09	00.10	&	-04	10	24.0	&	RRc		&	10.68	&	11.09	&	0.3032613	&	25	&	8.52	&	25\\
SZ Hya						&		&	09	13	48.68	&	-09	19	08.9	&	RRab	&	10.44	&	11.84	&	0.53724022&	3		&	26.23	&	23\\
RW Cnc						&	:	&	09	19	06.04	&	29	03	55.7	&	RRab	&	10.7	&	12.6	&	0.547199	&	3		&	87	&	26\\
UU Hya						&		&	09	36	29.73	&	04	06	40.3	&	RRab	&	11.73	&	12.73	&	0.5238684	&	3		&	39.89	&	23\\
ASAS093731-1816.2	&		&	09	37	31.00	&	-18	16	12.0	&	RRab	&	13.05	&	14.28	&	0.529175	&	25	&	87.73	&	25\\
CM UMa						&		&	09	43	13.78	&	49	29	37.3	&	RRab	&	12.8	&	13.8	&	0.589124	&	10	&	27.77	&	10\\
CD Vel						&		&	09	44	38.24	&	-45	52	37.2	&	RRab	&	11.3	&	12.4	&	0.5735076	&	25	&	66.35	&	25\\
ASAS101200+1921.9	&	:	&	10	12	00.00	&	19	21	54.0	&	RRab	&	11.55	&	12.43	&	0.4826394	&	25	&	1141.03	&	25\\
Y LMi							&	:	&	10	15	51.45	&	32	51	32.5	&	RRab	&	11.4	&	13.3	&	0.524471	&	3		&	33.4	&	26\\
Cze V134					&	:	&	10	22	26.27	&	59	12	36.2	&	RRc		&	11.46	&	11.63	&	0.419794	&	3		&	...	&	44\\
V543 Hya					&		&	10	26	08.40	&	-23	15	13.9	&	RRab	&	12.8	&	13.9	&	0.59826		&	17	&	59	&	17\\
BT Ant						&	:	&	10	32	02.59	&	-30	10	37.3	&	RRc		&	11.55	&	12.02	&	0.3304459	&	25	&	1731.6	&	25\\
AF Vel						&		&	10	53	02.49	&	-49	54	22.7	&	RRab	&	10.68	&	11.78	&	0.5274139	&	25	&	58.55	&	23\\
SZ Leo						&		&	11	01	36.82	&	08	09	55.6	&	RRab	&	11.79	&	12.72	&	0.53408		&	17	&	179	&	17\\
AH Leo						&	:	&	11	05	05.29	&	23	21	08.4	&	RRab	&	13.67	&	14.66	&	0.4662609	&	1		&	$\sim$20	&	1\\
ASAS110522-2641.0	&		&	11	05	22.00	&	-26	41	00.0	&	RRc		&	11.68	&	12.06	&	0.2944559	&	25	&	7.4	&	25\\
ASAS112027-4338.8	&	:	&	11	20	26.70	&	-43	38	48.0	&	RRc		&	11.05	&	11.34	&	0.3795948	&	25	&	1480.38	&	25\\
AI Crt						&	:	&	11	26	07.49	&	-14	03	43.1	&	RRab	&	15.5	&	16.1	&	0.50290		&	17	&	63	&	17\\
V354 Vir					&	:	&	11	43	32.22	&	02	41	55.6	&	RRab	&	12.4	&	13.3	&	0.59503		&	3		&	59	&	27\\
BI Cen						&		&	11	45	54.65	&	-59	22	40.2	&	RRab	&	11.18	&	12.33	&	0.4531949	&	3		&	79.68	&	23\\
X Crt							&		&	11	48	56.22	&	-10	26	28.6	&	RRab	&	11.09	&	11.75	&	0.73284		&	17	&	143	&	17\\
IK Hya						&	:	&	12	04	47.27	&	-27	40	43.3	&	RRab	&	9.96	&	10.42	&	0.653243	&	25	&	72	&	17\\
									&		&								&								&				&				&				&						&			&	67.5&	23\\
EL Hya						&	:	&	12	09	42.07	&	-34	57	26.3	&	RRc		&	13.3	&	13.8	&	0.3436271	&	25	&	1611.6	&	25\\
V552 Hya					&		&	12	12	06.12	&	-26	12	48.2	&	RRab	&	12.5	&	13.8	&	0.39878		&	17	&	48.3	&	17\\
TU Com						&	:	&	12	13	46.92	&	30	59	07.5	&	RRab	&	12.9	&	16.4	&	0.461809	&	3		&	$\sim$75.00	&	3\\
SV Hya						&		&	12	30	30.50	&	-26	02	51.1	&	RRab	&	9.78	&	11.00	&	0.4785475	&	25	&	63.29	&	25\\
ASAS123812-4422.5	&	:	&	12	38	12.50	&	-44	22	30.0	&	RRab	&	13.1	&	13.82	&	0.523549	&	25	&	1307.7	&	25\\
V419 Vir					&		&	12	48	04.51	&	-08	20	47.4	&	RRab	&	11.98	&	12.86	&	0.515287	&	25	&	65.69	&	25\\
Z CVn							&		&	12	49	45.42	&	43	46	25.6	&	RRab	&	11.43	&	12.36	&	0.653819	&	3		&	22.98	&	23\\
AS Vir						&	:	&	12	52	45.86	&	-10	15	36.4	&	RRab	&	11.60	&	12.23	&	0.55340399&	3		&	...	&	34\\
RY Com						&		&	13	05	07.99	&	23	16	42.2	&	RRab	&	11.75	&	12.92	&	0.468951	&	8		&	32	&	8\\
UZ Vir						&		&	13	08	44.34	&	13	24	08.3	&	RRab	&	12.58	&	13.70	&	0.4593925	&	9		&	68.24	&	9\\
OR Com						&		&	13	19	54.50	&	19	53	36.8	&	RRab	&	12.6	&	13.7	&	0.601167	&	3		&	74	&	27\\
AM Vir						&	:	&	13	23	33.33	&	-16	39	57.9	&	RRab	&	11.16	&	11.85	&	0.61509		&	17	&	49.8	&	17\\
V476 Vir					&	:	&	13	29	22.48	&	-05	52	59.2	&	RRab	&	11.2	&	12.1	&	0.5763497	&	25	&	1398.99	&	25\\
SS CVn						&		&	13	48	15.94	&	39	54	03.0	&	RRab	&	11.53	&	12.62	&	0.47851		&	3		&	93.72	&	23\\
ASAS135740-1202.3	&	:	&	13	57	40.00	&	-12	02	18.0	&	RRc		&	12.30	&	12.71	&	0.2671226	&	25	&	1189.91	&	25\\
ASAS135813-4215.1	&		&	13	58	13.00	&	-42	15	06.0	&	RRab	&	12.43	&	13.27	&	0.5231816	&	25	&	146.01	&	25\\
V674 Cen					&	:	&	14	03	24.08	&	-36	24	20.1	&	RRab	&	11.0	&	11.9	&	0.4939666	&	25	&	1650.71	&	25\\
ASAS141025-2244.8	&	:	&	14	10	25.00	&	-22	44	48.0	&	RRab	&	12.60	&	13.29	&	0.6398808	&	25	&	1556.66	&	25\\
V559 Hya					&		&	14	13	45.50	&	-22	54	41.9	&	RRab	&	11.9	&	13.1	&	0.44794		&	17	&	26.6	&	17\\
TV Boo						&		&	14	16	36.58	&	42	21	35.7	&	RRc		&	10.71	&	11.30	&	0.31255936&	3		&	10	&	27\\
V551 Vir					&		&	14	23	05.58	&	01	54	00.9	&	RRab	&	12.7	&	13.8	&	0.44692		&	3		&	48	&	27\\
SW Boo						&	:	&	14	27	34.86	&	36	02	44.1	&	RRab	&	11.76	&	12.88	&	0.5135281	&	3		&	13	&	26\\
ST Vir						&		&	14	27	39.08	&	00	54	05.8	&	RRab	&	10.84	&	12.15	&	0.4108143	&	3		&	25.58	&	23\\
RS Boo						&		&	14	33	33.21	&	31	45	16.6 	&	RRab	&	9.69	&	10.84	&	0.37733896&	3		&	532.48	&	23\\
ASAS144154-0324.7	&		&	14	41	54.00	&	-03	24	42.0	&	RRc		&	11.40	&	11.72	&	0.2293674	&	25	&	5.65	&	25\\
TY Aps						&		&	14	48	50.01	&	-71	19	41.9	&	RRab	&	11.20	&	12.43	&	0.5016935	&	3		&	109.13	&	23\\
MR Lib						&		&	14	53	15.44	&	-14	35	56.8	&	RRab	&	12.4	&	13.4	&	0.54007		&	17	&	41.7	&	17\\
V339 Lup					&		&	15	03	27.43	&	-47	56	03.7	&	RRab	&	11.5	&	12.3	&	0.60058		&	17	&	59.5	&	17\\
FU Lup						&		&	15	09	23.77	&	-43	19	37.1	&	RRab	&	14.0	&	15.0	&	0.3821508	&	25	&	42.49	&	25\\
ASAS151849-1000.0	&		&	15	18	49.00	&	-10	00	00.0	&	RRc		&	12.03	&	12.53	&	0.3364272	&	25	&	802.95	&	25\\
ST Boo						&		&	15	30	39.23	&	35	47	04.3	&	RRab	&	10.49	&	11.41	&	0.62229069&	3		&	284.09	&	23\\
AR Ser						&	:	&	15	33	30.82	&	02	46	37.9	&	RRab	&	11.6	&	12.2	&	0.5752124	&	3		&	63	&	27\\
CG Lib						&	:	&	15	35	16.81	&	-24	20	12.5	&	RRc		&	11.2	&	11.8	&	0.306777	&	25	&	1560.06	&	25\\
ASAS153830-6906.4	&		&	15	38	30.00	&	-69	06	24.0	&	RRab	&	12.26	&	13.15	&	0.6224747	&	25	&	118.05	&	25\\
V1141 Her					&	:	&	15	54	58.55	&	42	46	10.5	&	RRc		&	10.97	&	11.56	&	0.317152	&	3		&	$\sim$30	&	45\\
V1319 Sco					&	:	&	15	55	51.59	&	-21	48	32.9	&	RRc		&	11.35	&	11.90	&	0.2541338	&	25	&	1699.52	&	25\\
PQ Lup						&		&	15	55	53.25	&	-40	41	43.6	&	RRab	&	11.6	&	12.4	&	0.58198		&	17	&	48.8	&	17\\
AR Her						&		&	16	00	32.23	&	46	55	25.7	&	RRab	&	10.59	&	11.63	&	0.470028	&	3		&	32	&	27\\
BT Sco						&		&	16	12	55.58	&	-08	27	28.0	&	RRab	&	12.61	&	13.40	&	0.54871		&	17	&	78	&	17\\
GSC02050-00745		&		&	16	18	34.34	&	27	28	13.2	&	RRab	&	14.27	&	...		&	0.508646	&	3		&	...	&	36\\
BS Aps						&		&	16	20	51.50	&	-71	40	15.8	&	RRab	&	11.85	&	12.49	&	0.5825589	&	3		&	40.93	&	23\\
ASAS162158+0244.5	&		&	16	21	58.00	&	02	44	30.0	&	RRc		&	12.47	&	12.99	&	0.3238044	&	25	&	8.11	&	25\\
ASAS162811+0304.3	&		&	16	28	11.00	&	03	04	18.0	&	RRab	&	13.09	&	14.74	&	0.5970104	&	25	&	26.28	&	25\\
VX Her						&		&	16	30	40.80	&	18	22	00.6	&	RRab	&	9.91	&	11.18	&	0.4553573	&	3		&	455.37	&	39\\
UV Oct						&	:	&	16	32	25.53	&	-83	54	10.5	&	RRab	&	8.70	&	9.97	&	0.542625	&	3		&	143.73	&	25\\
									&		&								&								&				&				&				&						&			&	145	&	17\\
									&		&								&								&				&				&				&						&			&	146.99	&	23\\
RW Dra						&		&	16	35	31.60	&	57	50	23.2	&	RRab	&	11.05	&	12.08	&	0.442917	&	3		&	41.42	&	23\\
ASAS170223-2422.0	&		&	17	02	23.18	&	-24	21	59.2	&	RRab	&	11.34	&	11.73	&	0.4613693	&	25	&	22.18	&	25\\
V1124 Her					&		&	17	04	32.90	&	14	26	33.0	&	RRab	&	12.10	&	12.95	&	0.55102		&	27	&	39	&	27\\
V365 Her					&	:	&	17	05	39.86	&	21	30	58.0	&	RRab	&	12.61	&	13.55	&	0.3797141	&	3		&	40	&	27\\
DL Her						&		&	17	20	22.45	&	14	30	38.7	&	RRab	&	11.72	&	12.63	&	0.5916369	&	3		&	34	&	27\\
ASAS172721-5305.9	&		&	17	27	21.00	&	-53	05	54.0	&	RRab	&	12.30	&	13.55	&	0.435433	&	25	&	58.66	&	25\\
EZ Ara						&	:	&	17	29	31.78	&	-55	48	18.7	&	RRc		&	12.9	&	13.7	&	0.3273052	&	25	&	1610.05	&	25\\
V421 Her					&	:	&	17	32	05.47	&	39	45	32.2	&	RRab	&	13.33	&	14.46	&	0.55677		&	3		&	56	&	27\\
V788 Oph					&	:	&	17	36	09.07	&	08	09	54.1	&	RRab	&	13.3	&	14.9	&	0.547131	&	3		&	115	&	3\\
V434 Her					&	:	&	17	40	33.01	&	22	49	02.3	&	RRab	&	13.8	&	15.2	&	0.5144034	&	3		&	26.1	&	26\\
V494 Sco					&	:	&	17	40	48.48	&	-31	32	31.8	&	RRab	&	10.62	&	11.91	&	0.427297	&	25	&	455	&	17\\
									&		&								&								&				&				&				&						&			&	504.03	&	25\\
ASAS174202-4633.7	&	:	&	17	42	02.00	&	-46	33	42.0	&	RRc		&	10.74	&	11.11	&	0.3115788	&	25	&	1706.78	&	25\\
V829 Oph					&	:	&	17	49	29.20	&	12	13	54.5	&	RRab	&	13.5	&	15.0	&	0.56923		&	3		&	165	&	3\\
S Ara							&		&	17	59	10.73	&	-49	26	00.5	&	RRab	&	9.92	&	11.24	&	0.45186		&	17	&	49.37	&	25\\
AV Dra						&		&	17	59	44.21	&	51	53	01.7	&	RRab	&	12.50	&	13.57	&	0.55560		&	3		&	96	&	27\\
ASAS180023-7026.5	&	:	&	18	00	23.00	&	-70	26	30.0	&	RRc		&	12.08	&	12.47	&	0.3556146	&	25	&	1162.79	&	25\\
WW CrA						&		&	18	05	36.79	&	-43	49	57.4	&	RRab	&	11.66	&	12.48	&	0.55949		&	17	&	35.5	&	17\\
ASAS181215-5206.9	&		&	18	12	15.00	&	-52	06	54.0	&	RRab	&	12.59	&	13.26	&	0.8375462	&	25	&	5.22	&	25\\
V442 Her					&	:	&	18	12	58.32	&	42	03	45.5	&	RRab	&	12.5	&	13.8	&	0.442084	&	3		&	$\geq$700.00	&	29\\
BD Dra						&		&	18	17	51.94	&	77	17	49.2	&	RRab	&	12.05	&	13.01	&	0.58902	&	3		&	24.11	&	23\\
MW Lyr						&		&	18	19	53.82	&	31	58	54.6	&	RRab	&	12.5	&	14.0	&	0.3976742	&	5		&	16.55	&	5\\
ASAS182913+2104.3	&		&	18	29	13.00	&	21	04	18.0	&	RRab	&	11.26	&	12.18	&	0.371117	&	3		&	23	&	27\\
KM Lyr						&	:	&	18	30	29.76	&	40	18	15.8	&	RRab	&	12.8	&	13.9	&	0.500193	&	3		&	30	&	26\\
KX Lyr						&	:	&	18	33	15.22	&	40	10	22.8	&	RRab	&	10.38	&	11.47	&	0.44090446&	3		&	...	&	41\\
BH Pav						&		&	18	34	40.57	&	-65	27	03.0	&	RRab	&	11.5	&	13.1	&	0.4769536	&	25	&	173.7	&	25\\
AQ Lyr						&		&	18	34	51.04	&	26	35	41.8	&	RRab	&	12.30	&	13.51	&	0.357134	&	8		&	64.9	&	8\\
CoRoT105288363		&		&	18	39	30.86	&	07	26	53.6	&	RRab	&	14.96	&	15.66	&	0.56744122&	3		&	35.6	&	32\\
V413 CrA					&		&	18	47	57.62	&	-37	44	22.5	&	RRab	&	10.23	&	10.90	&	0.5893445	&	25	&	59.96	&	25\\
V349 Lyr					&	:	&	18	49	24.27	&	42	44	45.2	&	RRab	&	16.78	&	17.77	&	0.507074	&	31	&	$\geq$127	&	31\\
BD Her						&		&	18	50	32.19	&	16	31	50.9	&	RRab	&	11.72	&	12.63	&	0.4739064	&	3		&	$\sim$22	&	8\\
V353 Lyr					&		&	18	52	01.78	&	45	18	31.4	&	RRab	&	16.0	&	17.0	&	0.55682		&	31	&	60	&	31\\
V354 Lyr					&	:	&	18	52	50.27	&	41	33	49.4	&	RRab	&	15.0	&	16.0	&	0.56168		&	31	&	$\geq$127	&	31\\
V355 Lyr					&		&	18	53	25.83	&	43	09	16.2	&	RRab	&	13.8	&	15.3	&	0.473697	&	31	&	31.4	&	31\\
ASAS185719-6321.4	&		&	18	57	19.00	&	-63	21	24.0	&	RRab	&	12.28	&	13.24	&	0.41217		&	25	&	61.39	&	25\\
KIC11125706				&		&	19	00	58.78	&	48	44	41.6	&	RRab	&	11.83	&	12.26	&	0.61324		&	31	&	39.4	&	31\\
V360 Lyr					&		&	19	01	58.53	&	46	26	45.7	&	RRab	&	15.5	&	16.5	&	0.55759		&	31	&	51.4	&	31\\
NR Lyr						&	:	&	19	08	27.26	&	38	48	46.0	&	RRab	&	12.22	&	12.98	&	0.6820264	&	3		&	27	&	27\\
V450 Lyr					&	:	&	19	09	36.66	&	43	21	50.0	&	RRab	&	14.3	&	16.7	&	0.50461		&	31	&	$\sim$125	&	31\\
V366 Lyr					&		&	19	09	40.65	&	46	17	18.1	&	RRab	&	15.5	&	16.5	&	0.52702		&	31	&	65.6	&	31\\
V1104 Cyg					&		&	19	18	00.41	&	50	45	17.5	&	RRab	&	14.5	&	15.5	&	0.43639		&	31	&	53.1	&	31\\
V1127 Aql					&		&	19	24	00.11	&	01	41	48.9	&	RRab	&	14.8	&	16.0	&	0.355997	&	42	&	26.88	&	42\\
CoRoT100881648		&		&	19	25	05.43	&	01	39	23.8	&	RRab	&	14.94	&	...		&	0.60700		&	23	&	59.8	&	23\\
CoRoT101128793		&		&	19	26	37.32	&	01	13	34.9	&	RRab	&	15.93	&	16.53	&	0.4719296	&	3		&	18	&	33\\
ASAS192824-1852.4	&	:	&	19	28	24.00	&	-18	52	24.0	&	RRc		&	12.65	&	13.10	&	0.3563567	&	25	&	1572.33	&	25\\
CoRoT101503544		&		&	19	29	10.13	&	00	43	46.9	&	RRab	&	14.52	&	...		&	0.60500		&	23	&	25.6	&	23\\
WY Dra						&		&	19	33	20.76	&	80	55	42.9	&	RRab	&	12.08	&	13.64	&	0.588941	&	3		&	14.3	&	28\\
ASAS193538-7409.9	&	:	&	19	35	38.00	&	-74	09	55.0	&	RRc		&	12.57	&	13.08	&	0.3499993	&	25	&	1608.49	&	25\\
V2178 Cyg					&	:	&	19	40	06.99	&	38	58	20.4	&	RRab	&	15.5	&	17.0	&	0.48680		&	31	&	$\geq$200	&	31\\
ASAS194502+2434.2	&		&	19	45	02.00	&	24	34	12.0	&	RRab	&	11.72	&	12.02	&	0.8458661	&	25	&	37.56	&	25\\
V808 Cyg					&	:	&	19	45	39.07	&	39	30	54.8	&	RRab	&	15.3	&	16.6	&	0.5478641	&	31	&	$\sim$90	&	31\\
FO Pav						&		&	19	51	42.20	&	-62	44	07.8	&	RRab	&	11.3	&	12.2	&	0.5514395	&	25	&	557.17	&	25\\
V783 Cyg					&		&	19	52	52.71	&	40	47	35.4	&	RRab	&	14.2	&	15.5	&	0.6206994	&	31	&	27.7	&	31\\
ASAS195927-3400.1	&		&	19	59	26.70	&	-34	00	03.5	&	RRab	&	11.88	&	12.65	&	0.37972		&	17	&	45.7	&	17\\
V759 Cyg					&		&	20	00	26.80	&	48	59	37.8	&	RRab	&	12.1	&	13.8	&	0.360014	&	8		&	16	&	8\\
ASAS200431-5352.3	&		&	20	04	31.40	&	-53	52	20.0	&	RRc		&	10.95	&	12.26	&	0.32402		&	25	&	10.82	&	25\\
KM Aql						&		&	20	05	56.27	&	-08	30	52.4	&	RRab	&	12.7	&	13.8	&	0.4381966	&	25	&	192.2	&	25\\
V2239 Sgr					&		&	20	09	09.67	&	-41	49	31.8	&	RRab	&	12.0	&	13.0	&	0.441943	&	25	&	45.39	&	25\\
V1645 Sgr					&	:	&	20	20	44.47	&	-41	07	05.7	&	RRab	&	11.5	&	12.1	&	0.5529452	&	3		&	1331.74	&	25\\
GZ Del						&	:	&	20	22	24.53	&	10	34	07.3	&	RRab	&	15.4	&	16.7	&	0.33582841&	3		&	$\sim$36	&	38\\
ASAS202746-2850.5	&	:	&	20	27	45.70	&	-28	50	33.0	&	RRab	&	12.24	&	12.64	&	0.4084525	&	25	&	1674.48	&	25\\
ASAS203145-2158.7	&		&	20	31	45.00	&	-21	58	42.0	&	RRc		&	11.25	&	11.64	&	0.317152	&	25	&	792.83	&	25\\
ASAS203420-2508.9	&		&	20	34	20.00	&	-25	08	54.0	&	RRab	&	11.58	&	12.48	&	0.5262389	&	25	&	666.44	&	25\\
ASAS203749-5735.5	&	:	&	20	37	49.00	&	-57	35	30.0	&	RRc		&	12.38	&	12.66	&	0.4199162	&	25	&	1270.33	&	25\\
ASAS204440-2402.7	&		&	20	44	39.90	&	-24	02	44.0	&	RRc		&	12.75	&	13.11	&	0.205333	&	25	&	6.64	&	25\\
FK Vul						&		&	20	52	31.00	&	22	26	11.7	&	RRab	&	12.06	&	12.95	&	0.4340527	&	3		&	56	&	8\\
RV Cap						&		&	21	01	28.87	&	-15	13	46.1	&	RRab	&	10.22	&	11.57	&	0.4477465	&	25	&	231.66	&	25\\
ASAS210741-5844.2	&	:	&	21	07	41.00	&	-58	44	12.0	&	RRc		&	13.33	&	13.78	&	0.3462376	&	25	&	1479.95	&	25\\
Z Mic							&	:	&	21	16	22.71	&	-30	17	03.1	&	RRab	&	11.26	&	11.92	&	0.5869258	&	3		&	...	&	34\\
ASAS211839+0612.3	&	:	&	21	18	39.40	&	06	12	18.0	&	RRc		&	11.04	&	11.53	&	0.2914601	&	25	&	1176.75	&	25\\
ASAS212034+1837.2	&		&	21	20	34.00	&	18	37	12.0	&	RRab	&	11.50	&	12.26	&	0.5624065	&	25	&	81.3	&	25\\
DM Cyg						&		&	21	21	11.55	&	32	11	28.7	&	RRab	&	10.93	&	11.99	&	0.419863	&	7		&	10.57	&	7\\
ASAS212331-3025.0	&	:	&	21	23	31.00	&	-30	25	00.0	&	RRc		&	12.35	&	12.84	&	0.367442	&	25	&	1739.74	&	25\\
ASAS212433-5712.1	&		&	21	24	33.24	&	-57	12	04.2	&	RRab	&	12.95	&	13.99	&	0.6051401	&	17	&	133.38	&	25\\
RY Oct						&		&	21	36	09.37	&	-77	18	13.5	&	RRab	&	11.46	&	12.46	&	0.563469	&	3		&	216.45	&	23\\
ASAS213826-3945.0	&	:	&	21	38	26.00	&	-39	44	57.0	&	RRc		&	13.07	&	13.56	&	0.4107031	&	25	&	1540.12	&	25\\
ASAS214101+0109.6	&		&	21	41	01.00	&	01	09	36.0	&	RRab	&	12.43	&	13.0	&	0.6156709	&	25	&	522.58	&	25\\
RS Oct						&		&	21	47	16.91	&	-87	39	06.4	&	RRab	&	12.2	&	13.4	&	0.458038	&	3		&	244.2	&	25\\
RT Gru						&		&	21	51	58.44	&	-45	59	06.7	&	RRab	&	12.19	&	13.15	&	0.51216		&	17	&	87	&	17\\
SS Oct						&		&	21	53	35.38	&	-82	46	43.8	&	RRab	&	10.8	&	12.1	&	0.6218493	&	25	&	144.93	&	23\\
BV Aqr						&	:	&	22	02	54.00	&	-21	31	32.1	&	RRc		&	10.72	&	11.24	&	0.363714	&	3		&	11.2	&	26\\
									&		&								&								&				&				&				&						&			&	1413.43	&	25\\
ASAS221556-2522.6	&		&	22	15	55.70	&	-25	22	39.0	&	RRab	&	11.30	&	12.08	&	0.5467383	&	25	&	5.78	&	25\\
TY Gru						&	:	&	22	16	39.42	&	-39	56	18.0	&	RRab	&	13.6	&	14.7	&	0.570076	&	3		&	...	&	34\\
GP Aqr						&	:	&	22	25	39.14	&	-07	56	27.7	&	RRc		&	10.66	&	11.04	&	0.4052637	&	25	&	1618.65	&	25\\
AE Peg						&	:	&	22	27	21.54	&	16	48	16.7	&	RRab	&	11.83	&	13.15	&	0.4967235	&	3		&	23	&	27\\
ASAS223427-5635.4	&		&	22	34	27.00	&	-56	35	24.0	&	RRab	&	13.07	&	13.45	&	0.61499		&	17	&	63	&	17\\
ASAS225131-3006.2	&	:	&	22	51	31.00	&	-30	06	12.0	&	RRc		&	13.21	&	13.66	&	0.3384769	&	25	&	1681.8	&	25\\
ASAS225248-2442.2	&		&	22	52	47.80	&	-24	42	12.0	&	RRab	&	12.78	&	13.97	&	0.5295565	&	25	&	181.2	&	25\\
BH Peg						&	:	&	22	53	01.04	&	15	47	16.6	&	RRab	&	9.99	&	10.79	&	0.640993	&	3		&	39.8	&	26\\
ASAS225323+0846.1	&		&	22	53	23.20	&	08	46	09.0	&	RRab	&	12.60	&	14.11	&	0.4930493	&	25	&	348.58	&	25\\
ASAS225518-2317.6	&	:	&	22	55	18.00	&	-23	17	36.0	&	RRc		&	13.00	&	13.55	&	0.3935794	&	25	&	1557.88	&	25\\
BO Gru						&		&	23	06	58.64	&	-43	54	38.5	&	RRc		&	12.1	&	12.6	&	0.2811062	&	25	&	10.24	&	25\\
ASAS231209-1855.4	&	:	&	23	12	09.00	&	-18	55	24.0	&	RRc		&	12.73	&	13.05	&	0.3079943	&	25	&	1349.89	&	25\\
GV And						&	:	&	23	13	12.56	&	36	54	04.0	&	RRab	&	13.07	&	14.03	&	0.528092	&	16	&	$\sim$32	&	16\\
ASAS232031-1447.9	&		&	23	20	31.00	&	-14	47	54.0	&	RRab	&	12.46	&	13.05	&	0.6269552	&	25	&	54.52	&	25\\
ASAS233951-1644.4	&		&	23	39	51.30	&	-16	44	25.0	&	RRc		&	12.16	&	12.68	&	0.3553741	&	25	&	875.96	&	25\\
DY And						&	:	&	23	58	42.21	&	41	29	19.4	&	RRab	&	12.94	&	14.15	&	0.6030897	&	3		&	...	&	35\\

\end{longtable}
\end{footnotesize}

\begin{table*}[htbp]
\centering
\caption{Blazhko stars with multiple modulations}
\begin{tabular}{lccrlrrcrrrr}
\hline 
\hline
		Star	& &	RA	&	DE~~~~~~	&	Type	&	V$_{\mathrm{max}}$	&	V$_{\mathrm{min}}$ 	&	P$_{\mathrm{puls}}$	&	Ref.	&	P$_{\mathrm{mod1}}$ & P$_{\mathrm{mod2}}$ 	&	Ref.	\\
					&	&			&						&				&	[mag]								&	[mag]								&	[d]							&				&	[d]~~									&	[d]~~									&		\\
\hline
	SU	Col	&:\tablefootmark{a}&	05 07 47.05	&	-33 51 54.5	&	RRab	&	11.32	&	13.33	&	0.4873552	&	25	&	65.41&88.98	&	25	\\
	UZ	UMa	&	&	08 18 53.94	&	+73 05 47.8	&	RRab	&	13.10	&	15.00	&	0.4668413	&	6		&	26.7  & 143     		&	8		\\
	LS	Her	&	&	16 02 03.79	&	+17 28 50.4	&	RRc		&	11.04	&	11.53	&	0.230808	&	21	&	12.75 & 109   			&	21	\\
	V872 Oph&:&	17 55 17.81	&	+08 13 42.9	&	RRab	&	14.70	&	15.80	&	0.45197319&	3		&	13.5  & 51.13				&	40	\\
	RZ	Lyr	&	&	18 43 37.88	&	+32 47 54.0	&	RRab	&	10.60	&	12.03	&	0.511230	&	11	&	121   & 30   				&	11	\\
	V445 Lyr& &	18 58 25.59	&	+41 35 48.6	&	RRab	&	15.30	&	17.30	&	0.513075	&	31	&	53.2	& 143.3       &	31	\\	
	XZ	Cyg	&	&	19 32 29.31	&	+56 23 17.5	&	RRab	&	8.90	&	10.16	&	0.46659934&	20	&	57.5  & 41.6				&	20	\\
	CZ	Lac	&	&	22 19 30.76	&	+51 28 14.8	&	RRab	&	10.77	&	11.26	&	0.432174	&	30	&	14.6  & 18.6				&	30	\\
\hline
\end{tabular}

\tablefoottext{a}{SU Col has one aditional modulation period 29.5 d}
\end{table*}

\begin{table*}[htbp]
\caption{Stars with changing Blazhko period}
\centering
\begin{tabular}{lccrlrrcrrrr}
\hline\hline
		Star	&	&	RA	&	DE~~~~~~	&	Type	&	V$_{\mathrm{max}}$	&	V$_{\mathrm{min}}$ 	&	P$_{\mathrm{puls}}$	&	Ref.	&	P$_{\mathrm{BLmin}}$ & P$_{\mathrm{BLmax}}$ 	&	Ref.	\\
					&	&			&						&				&	[mag]								&	[mag]								&	[d]									&				&	[d]~~									&	[d]~~									&	\\
\hline
	AD	UMa	&:&	09 23 38.66	&	+55 46 33.2	&	RRab	&	15.0	&	16.3	&	0.548315	&	3		&	35    & 40  	&	26	\\
	RV	UMa	&	&	13 33 18.09	&	+53 59 14.6	&	RRab	&	9.81	&	11.30	&	0.468060	&	3		&	89.9  & 90.63	&	12	\\
	XZ	Dra	&	&	19 09 42.61	&	+64 51 32.1	&	RRab	&	9.59	&	10.65	&	0.4764955	&	24	&	73    & 77  	&	24	\\
	RR 	Lyr	&	&	19 25 27.91	&	+42 47 03.7	&	RRab	&	7.06	&	8.12	&	0.566839	&	22	&	38.8  & 40.8	&	22	\\

\hline
\end{tabular}
\tablebib{(1) \citet{phillips2004}, (2) \citet{jurcsik2006}, (3) \citet{samus2012} or \citet{vsx}, (4) \citet{jurcsik2005}, (5) \citet{jurcsik2008}, (6) \citet{sodor2006}, (7) \citet{jurcsik2009b}, (8) \citet{jurcsik2009a}, (9) \citet{sodor2012}, (10) \citet{szeidl2012}, (11) \citet{jurcsik2012}, (12) \citet{hurta2008}, (13) \citet{achtenberg2001}, (14) \citet{lee2001b}, (15) \citet{lee2001a}, (16) \citet{lee2002}, (17) \citet{sodor2005}, (18) \citet{kovacs2005}, (19) \citet{kolenberg2009}, (20) \citet{lacluyze2004}, (21) \citet{wils2008}, (22) \citet{kolenberg2006}, (23) \citet{leborgne2012}, (24) \citet{jurcsik2002}, (25) \citet{szczygiel2007}, (26) \citet{smith1995}, (27) \citet{wils2006}, (28) \citet{chris1975}, (29) \citet{schmidt2000}, (30) \citet{sodor2011}, (31) \citet{benko2010}, (32) \citet{chadid2011}, (33) \citet{poretti2010}, (34) \citet{for2011}, (35) \citet{cano2010}, (36) \citet{antipin2005}, (37) \citet{oppenheim1998}, (38) \citet{hacke1988a}, (39) \citet{wunder1990}, (40) \citet{hacke1988b}, (41) \citet{firmanyuk1974}, (42) \citet{chadid2010}, 
(43) \citet{virnina2012}, (44) \citet{antipin2010}.
}

\end{table*}

\end{document}